# ROOM TEMPERATURE MAGNETOCALORIC EFFECT in $CrTe_{1-x}Se_x$ ALLOYS

M. Kh. Hamad[a*], I. C. Nlebedim[b], Yazan Maswadeh[c], R. Hamad[d], and Kh. A. Ziq[d§]

[a]*Department of Basic Sciences, School of Social and Basics Sciences, Al Hussein Technical University, Amman, Jordan*

[b]*Ames Laboratory, U.S. Department of Energy, Ames, IA 50011, United States.*

[c]*XRR/XRD Group, EAG Laboratory, Sunnyvale, CA 94086, United States.*

[d]*Physics Department, King Fahd University of Petroleum & Minerals (KFUPM), Dhahran, Saudi Arabia*

Corresponding authors: *morad.hamad@htu.edu.jo  §kaziq@kfupm.edu.sa

**Abstract**

In This work, we report room temperature magnetocaloric properties in *$CrTe_{1-x}Se_x$ (0.00≤x≤0.10)* alloys prepared by a conventional solid-state reaction. The Rietveld refinement of the XRD pattern of *$CrTe_{1-x}Se_x$* showed the emerging of pure hexagonal *NiAs* structure with $P6_3/mmc$ (194) space group with increasing Se substitution. Upon *$\mu_0 H$* = 5T applied magnetic field, the magnetocaloric effect analysis revealed maximum magnetic entropy change ($\Delta S_M$) in the range 7.9-9.2 J/kg.K with a relative cooling power (RCP) value of about 550-694 J/kg. The RCP values exceed the corresponding RCP value of the prototype Gd element near room temperature. The obtained results suggest that *$CrTe_{1-x}Se_x$ (0.00≤x≤0.10)* alloys is a viable candidate for a room temperature magnetic cooling application.

**Key words:** magnetocaloric effect; Relative cooling power; Rietveld refinement; Alloys

## 1- Introduction

Transition metals (M) Pnictogens/Chalcogens (X) materials have divers magnetic and electrical properties with potential practical applications in spintronic and magnetocaloric cooling among many other applications [1-2]. The binary compounds in the form of MX with broad solubility range in Pnictogens/Chalcogens substitution and commonly have nickel arsenide (NiAs) crystalline structure [3-4]. Recently, the Cr-based chalcogenides showed interesting van der Waals ferromagnetic properties, which revived wide research interest in reduced dimensional magnetic systems. For instance, Cr-monochalcogenides CrTe has hexagonal NiAs-type structure, reveals metallic behavior, which is ferromagnet with a Curie temperature $T_c \sim$ 320 to 340 K [5-9], and noncollinear magnetic state at $T_{cr} \sim$ 170 K [2]. The reported wide range in the $T_c$ is due to the non-stoichiometric effect inherent in this material [8-12]. Moreover, stoichiometric bulk CrTe does not exist in the pure hexagonal phase at room temperature [2, 8-9, 13-14]. On the other hand, the binary CrSe alloy is antiferromagnetic with $T_N \sim$ 300 K [2, 6, 13-15] with an effective spin magnetic moment of 4.9 $\mu_B$ [2]. The effects of Te/Se substitutions on the various physical properties of CrTe was previously investigated [2, 3, 6,11]. However, the magnetocaloric effects and critical behavior have not been investigated. It is worth mentioning that many materials with first order FM-PM phase transition found to have giant magnetocaloric effect. For example, the maximum change in entropy $|\Delta S_{max}|$ of $Ni_{37.5}Co_{12.5}Mn_{35}Ti_{15}$ is about 27 J/kg.K while it is about 6 J/kg.K for the pure gadolinium [16]. However, V. I. Zverev et al claims theoretically that a refrigerant significantly better than Gd cannot be found, and $\Delta T$ can never exceeds ~18 K/T [17]. V. I. Zverev and Radel R Gimaev concluded that, Rear Earth Metals (REM) and their alloys have high refrigeration capacity and are promising for use in magnetic refrigerators operating both in narrow temperature range (about 10–20K) and much wider temperature ranges. Zverev et al concluded that antiferromagnetic refrigeration based on REM and their alloys in the region of spiral structures existence have more advantage over ferromagnetic materials [18]. Other interesting applications of the magnetocaloric effect in medical applications is also discussed [19].

In the current study, we investigate the structural, and magnetocaloric effect for the $CrTe_{1-x}Se_x$ (0.00 ≤ x ≤ 0.20) system at room temperature. Our motivation is to find a material with a high relative cooling power that can be operated at and around room temperature over a wide range of temperature. We found that $CrTe_{1-x}Se_x$ system has an enhanced room temperature magnetocaloric effect compared to the pure gadolinium Gd, which is the prototype magnetocaloric material.

## 2- Experimental Techniques

Polycrystalline $CrTe_{1-x}Se_x$ samples were synthesized using the conventional solid-state reaction. Stoichiometric ratios of the high purity (>99.99%) elements (Cr, Te and Se) were mixed thoroughly in agate mortal then pressed into pellets. The pellets were sealed in evacuated quartz tubes partially filled with high-purity argon gas. The pellets were annealed at 800 ºC for 12 hours, then grinded, pressed, sealed, and reannealed at 1000 ºC for 24 hours. The samples were re-grinded, pressed into pellets and reannealed at 1000 ºC for another 24 hours. Powder x-ray diffraction (XRD) patterns were obtained using Bruker D2-Phaser diffractometer. We used FULLPROF software to analyze the diffraction patterns. Quantum Design MPMS3 SQUID magnetometer has been used to obtain the magnetization isotherms between 5 – 400 K.

### 3- Results and Discussions

#### A. *Structural analysis*

Figure 1 shows the XRD patterns fit using Rietveld refinement for $CrTe_{1-x}Se_x$ (0.0 ≤x≤ 0.1) samples. The results confirm two Cr-Te minority phases in all samples, monoclinic phase with $Cr_3Te_4$ and hexagonal phase with Cr:Te ratio of 1:1. The presence of the monoclinic phase structure at low Se-concentration (x ≤ 0.04) has been reported in earlier work [20]. The ICDD reference numbers with Wyckoff positions for the monoclinic and hexagonal phases are presented in Table 1. The hexagonal CrTe phase shows a preferred orientation along the z-axis, which is typical. The closed packed structure of the hexagonal CrTe phase gives an average Cr-Te bond length of 2.767Å, the monoclinic deformed version of the hexagonal structure that appears to have different stoichiometry with a relaxed structure shows slightly increased average Cr-Te bond length of 2.775Å. As Se concentration increases, the monoclinic phase decreases and essentially vanishes for x~10%. Table 2 shows the refined lattice constants, unit cell volume, and weight ratio of the hexagonal phase in the prepared samples at deferent Selenium concentrations. The overall decrease into the cell volume is between x = 0.02 – 0.10, which correlates with the decrease in the *a* and *b* lattice constants. The contraction of the unit cell is related to the smaller Se ionic radius compared to Te ionic radius. This is also reflected in the continuous shift of the Bragg peaks in Fig. 1 to lower 2θ angles. $CrTe_{0.90}Se_{0.10}$ which has the highest concentration of Selenium almost appears as a single hexagonal phase, with a trace amount of monoclinic phase weight of about 5%.

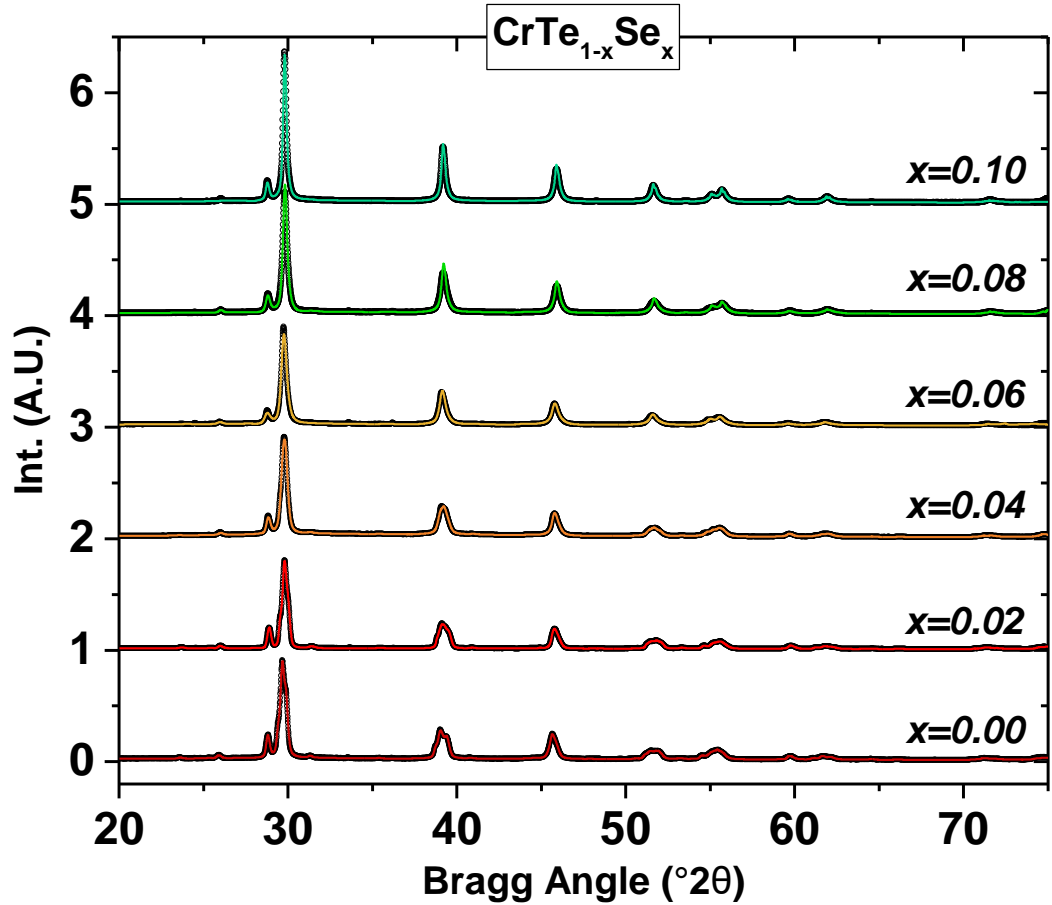

*Figure 1: XRD patterns (in black) with Rietveld refinements (in colors) for $CrTe_{1-x}Se_x$ (x= 0.00 to 0.10) sintered at 1000 °C.*

*Table 1: The ICDD reference numbers with Wyckoff positions for the monoclinic and hexagonal phases appeared in the samples.*

| [04-002-8473] $Cr_3Te_4$ Monoclinic C2/m (12) | | | | |
|---|---|---|---|---|
| Atom | Site | 1/a | 1/b | 1/c |
| Te | 4i | 0.1240 | 0.0000 | 0.4440 |
| Te | 4i | 0.3635 | 0.0000 | 0.0085 |
| Cr | 4i | 0.2630 | 0.0000 | 0.2750 |
| Cr | 2a | 0.0000 | 0.0000 | 0.0000 |
| | | | | |
| [04-002-6163] CrTe Hexagonal $P6_3/mmc$ (194) | | | | |
| Atom | Site | 1/a | 1/b | 1/c |
| Te | 2c | 0.3333 | 0.6667 | 0.2500 |
| Cr | 2a | 0.0000 | 0.0000 | 0.0000 |

*Table 2: Lattice constants and weight fraction of the hexagonal phase (P6₃/mmc) that obtained by XRD analysis of the CrTe$_{1-x}$Se$_x$ samples.*

| sample x | a=b (Å) | c (Å) | V (Å³) |
|---|---|---|---|
| 0.00 | 3.9649 | 6.1830 | 84.18 |
| 0.02 | 3.9668 | 6.2004 | 84.49 |
| 0.04 | 3.9643 | 6.2010 | 84.40 |
| 0.06 | 3.9594 | 6.2019 | 84.20 |
| 0.08 | 3.9594 | 6.2019 | 84.20 |
| 0.10 | 3.9561 | 6.2058 | 84.11 |

*B. Magnetic Properties*

Fig. 2 (a) shows the temperature dependences of magnetization *M(T)* for CrTe$_{1-x}$Se$_x$ (0.00≤x≤0.10) samples over the temperature range 5-350 K under an applied magnetic field of 0.05 T. The magnetization curves *M(T)* show that all the samples have a ferromagnetic to paramagnetic (FM-PM) phase transition near room temperature; $T_C$ decreased from 332 K (x=0.0 Se) to 295 K (x=0.1 Se) the data is presented in Table 3. The increasing Se contents may result in an increase of the disorder effect that is affecting the transition temperature [17]. The inflection points in the M(T) curves have been used to obtain the ferromagnetic transition temperature ($T_C$). The $T_C$ values are given in Table 3. It is worth mentioning that a wide variation in $T_C$ values was reported for the stoichiometric CrTe alloy. This is mainly due to difficulties in preparing a pure stoichiometric alloy. The $T_C$ of the unsubstituted sample (CrTe) reported here is 7 K higher than the value reported in [8] and [9]; however, it is in line with some other published works [10-12]. Similar behavior was observed in the three-dimensional CrTe$_{1-x}$Sb$_x$ [8].

The variations of the inverse DC-susceptibility with temperature are shown in Fig 2(b). In the paramagnetic region of FM-PM transition and at much higher temperatures than $T_C$; the magnetic susceptibility ($\chi$) follows the Curie-Weiss law, $\chi = C(T - \theta_c)^{-1}$, where C and $\theta_c$ are the Curie constant and the Curie temperature respectively. The linear extrapolation of the inverse susceptibility versus T gives the $\theta_c$. The obtained $T_C$ and $\theta_c$ values are given in Table 3. The values of $\theta_c$ are commonly higher than the $T_C$ values obtained from the inflection points of the *M(T)* curve suggesting a more complex spin configuration than normal FM-PM spin configuration. Similar results in CrSbSe$_3$ and in CrTe$_{1-x}$Sb$_x$ have been recently reported in [22, 8].

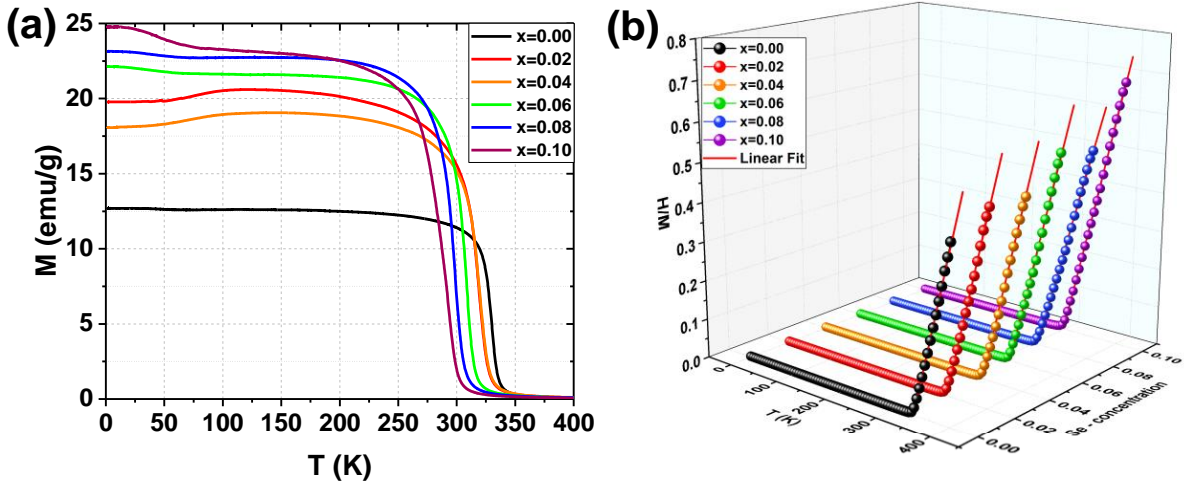

*Figure 2: (a) Variations of the magnetization ($\mu_0H= 0.05T$) with temperature for $CrTe_{1-x}Se_x$. (b) The temperature-dependent inverse susceptibility for different samples measured at 0.05 T.*

*Table 3: The Critical (Tc) and Curie ($\theta_c$) temperature for the $CrTe_{1-x}Se_x$ samples.*

| Se (x) | $T_C$ (K) | $\theta_c$ (K) |
|---|---|---|
| 0.00 | 332.(0) | 335.(3) |
| 0.02 | 320.(3) | 326.(6) |
| 0.04 | 318.(0) | 329.(0) |
| 0.06 | 310.(0) | 318.(1) |
| 0.08 | 299.(0) | 308.(3) |
| 0.10 | 295.(1) | 300.(0) |

Fig. 3(a) shows the isothermal magnetization curves for a selected sample ($CrTe_{0.90}Se_{0.10}$) at various temperatures, with $\Delta T=4K$. Below $T_C$, the sample is in its ferromagnetic state, with rapidly increasing magnetization (at low $\mu_0H\sim0.3T$), reaching saturation above 1T. Upon increasing temperature, the magnetization decreases monotonically, and the state moves toward the paramagnetic (PM) phase. In the PM region ( above $T_C$), the magnetization is essentially linear over the field range $\mu_0H = 0 - 5T$. Fig. 3(b) shows the isothermals curves for the studied $CrTe_{1-x}Se_x$ samples at 300 K. In the high field region (at T=300K) and as Se content increases, the magnetization at $\mu_0H = 5T$ gradually reduced, accompanied by a slight increase in high field susceptibility (slope). These changes are an indication of lowering the FM- exchange interaction resulting from the Se substitution for Te.

C. Magnetocaloric effect:

The changes in magnetic entropy ($\Delta S_M$) have been calculated from the magnetization isotherms (Fig. 3a) along with thermodynamic Maxwell's relations. The entropy change can be expressed as [23]:

$$\Delta S_M \left(\frac{T_1+T_2}{2}\right) = \frac{1}{T_1-T_2} \left[\int_0^{H_{max}} M(T_2, H)dH - \int_0^{H_{max}} M(T_1, H)dH\right] \quad (1)$$

Which is approximated at an average field and temperature as:

$$\Delta S_M(T,H) = \sum_i \frac{M_{i+1}(T_{i+1},H) - M_i(T_i,H)}{T_{i+1}-T_i} \Delta H \quad (2)$$

where $M_i$ and $M_{i+1}$ are the magnetization values measured in a field $H$, at temperature $T_i$ and $T_{i+1}$, respectively.

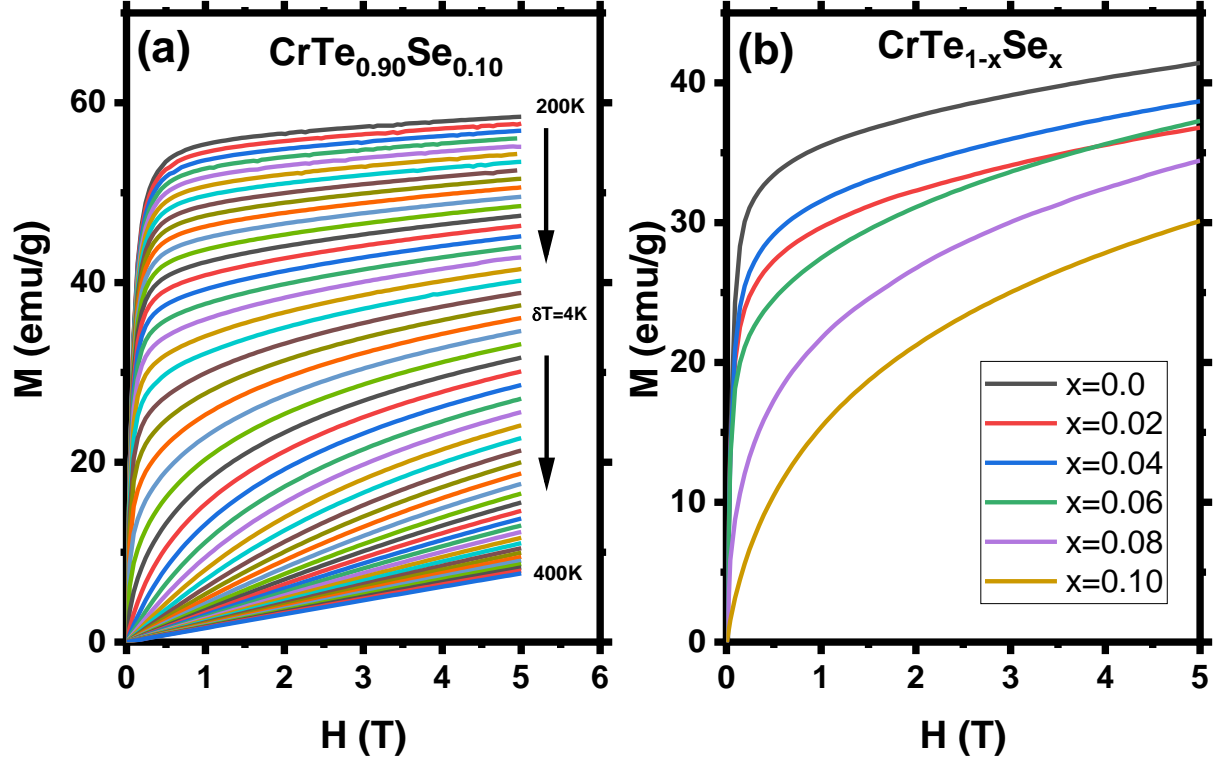

*Figure 3: Variations of the magnetization with magnetic field of: (a) CrTe$_{0.90}$Se$_{0.10}$ sample for ΔT=4K intervals. (b) Selected CrTe$_{1-x}$Se$_x$ samples at similar temperature (T=300 K).*

Fig. 4 shows the variation of [-ΔS(H,T)] as a function of temperature in the magnetic field range μ$_0$H = 0-5T for the studied samples. The change in magnetic entropy shows a peak ΔS$_M$ near the T$_C$ for all magnetic fields, which indicates that heat is released when the magnetic field is adiabatically changed [24]. The change in magnetic entropy varies with Se concentration, reaching a maximum value of 9.2 J/kg.K for CrTe$_{0.92}$Se$_{0.08}$. In the FM region (*i.e* T<T$_C$) the CrTe$_{1-x}$Se$_x$ samples show an oscillating behavior that decreases with decreasing H(max). These oscillations and completely vanished in the PM region (T>T$_C$). This behavior has been observed previously in several magnetocaloric materials and no conclusion has been reached about the cause or the origin of these oscillations [25, 26]. It is interesting to observe that the sample with no Se (CrTe) shows minimal oscillations. Moreover, for a given sample, these oscillations look similar at all investigated fields [27]. Figure 4 also reveals that the maximum magnetic entropy |-ΔS(H,T)| increases with the applied magnetic field *H* for all samples.

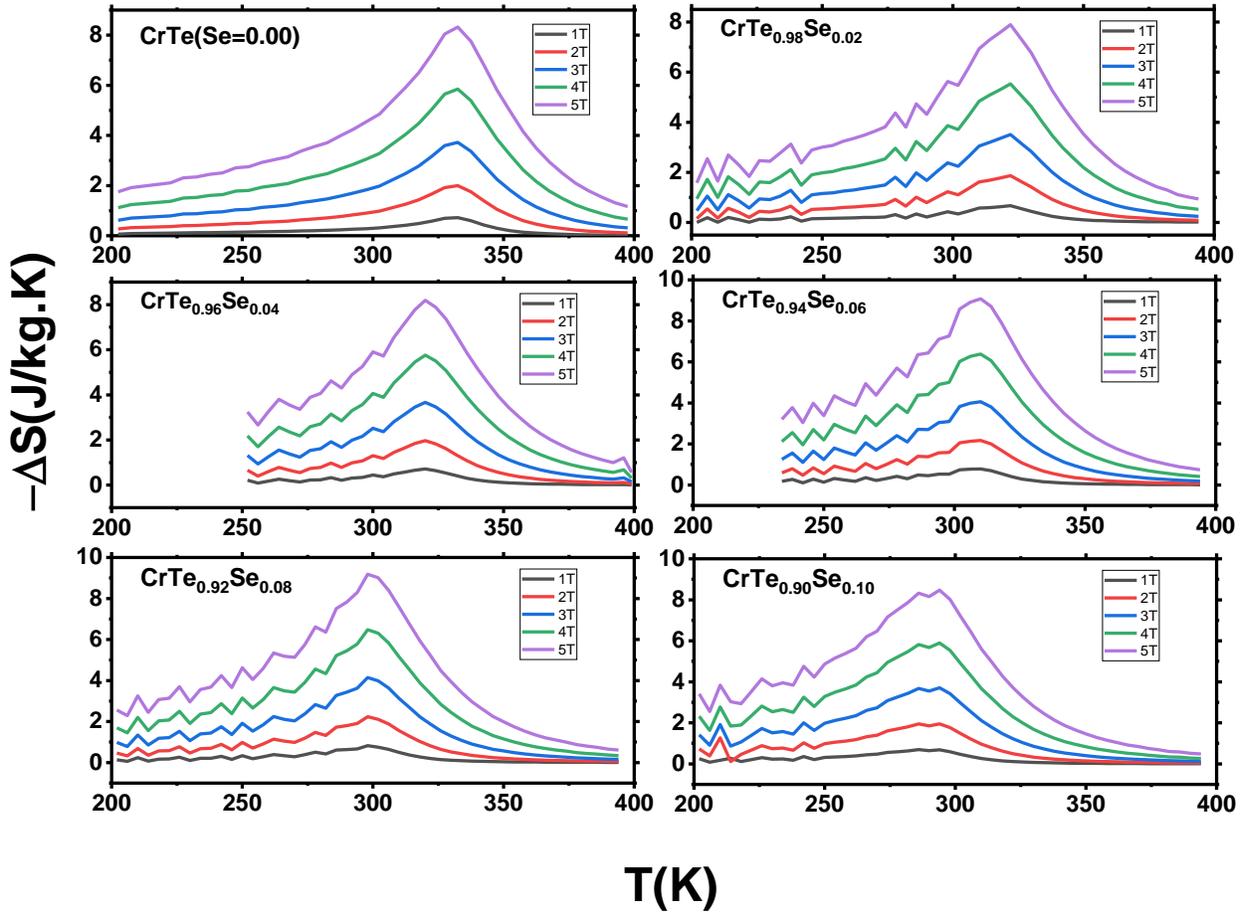

*Figure 4: Temperature dependence of |-ΔS(H,T)| of CrTe$_{1-x}$Se$_x$ at different cycling field.*

The relative cooling power RCP, is another important factor which is used to evaluate the cooling efficiency of magnetic cooling materials; it is defined as [28]:

$$RCP\ (S) = |\Delta S_M^{max}| \times \Delta T_{FWHM} \qquad (3)$$

Where, $\Delta S_M^{max}$ is the maximum change of magnetic entropy, and $\Delta T_{FWHM}$ is basically the practical cooling temperature range. The calculated RCP values for all CrTe$_{1-x}$Se$_x$ sample at different cycling magnetic fields are shown in Fig. 5. The RCP values closely follow a quadratic behavior with increasing magnetic field.

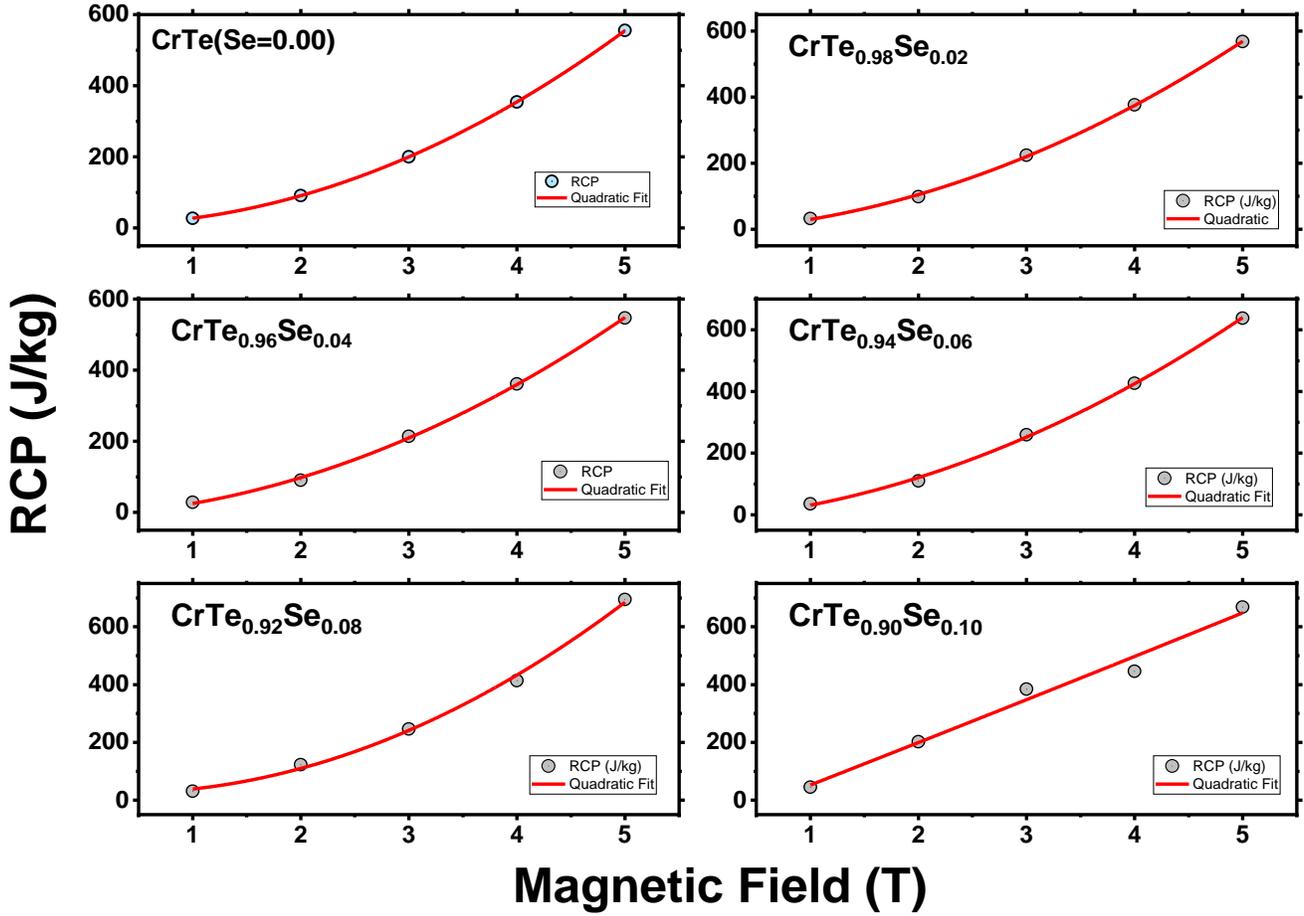

*Figure 5: Variation of the Relative cooling power (RCP) with the applied magnetic field ($\mu_0 H$) for the studied samples.*

To facilitate comparison to the cooling power of other materials, we present the change in magnetic entropy as a function of temperature at $\Delta\mu_0 H = 5T$ for all investigated samples in Fig. 6. It is easily seen that the peak position (the temperature) of the magnetic entropy can be systematically tuned by varying the Se contents. The figure also reveals that $|\text{-}\Delta S(H,T)|$ reaches maximum value (9.2 J/kg.K) for x = 0.08, which is also comparable with the value at x = 0.06 (9.1 J/kg.K). Both samples also have $T_C$ values near room temperature, 299 K and 310 K, respectively. The maximum values of $|\text{-}\Delta S(H,T)|$ for all studied samples are presented in Table 4 along with RCP values and for our samples, and the prototype magnetocaloric material's pure Gd and other materials obtained at

$\Delta\mu_0 H = 5$ Tesla [29-32]. It is noteworthy that the reported magnetic entropy values are larger than the Gd values at room temperature. All RCP values obtained in the present work are presented in Table 4, along with RCP value reported for $Gd_5Si_{2.06}Ge_{1.94}$. However, $Gd_5Si_{2.06}Ge_{1.94}$ contains nearly 80 wt.% of Gd, a critical rare earth element. Table 4 also shows $MnFeP_{0.45}As_{0.5}$ and two lanthanum-manganese perovskite compounds. The $T_C$ values of $MnFeP_{0.45}As_{0.5}$ is about 300K and for the two lanthanum-manganese perovskite compounds Tc is about 270K, however; their RCP values are substantially less than those for the $CrTe_{1-x}Se_x$ samples. Therefore, in addition to the high magnetic cooling efficiency, the $CrTe_{1-x}Se_x$ samples have $T_C$ values (295 – 320 K) in a range suitable for room temperature magnetic refrigeration application. Furthermore, the adiabatic width ($\Delta T_{ad}$) is approximately calculated according to [17]:

$$\Delta T_{ad} = -\left(\frac{T}{C_B}\right) \times \Delta S_M^{max} \tag{4}$$

Where T~300K and $C_B \sim 3R/mol$. The calculated $\Delta T_{ad}$ for x=0.08 and 0.06 samples, at $\mu_0 H = 1$ Tesla is approximately 1.7 K/T comparing to 3K/T for the pure Gd [17]. However, the operation range cooling is determined by the width of $|\Delta S_M^{max}|$ which is considerably larger than that in Gd. Table 4 shows the $\Delta T_{ad}$ for all studied samples.

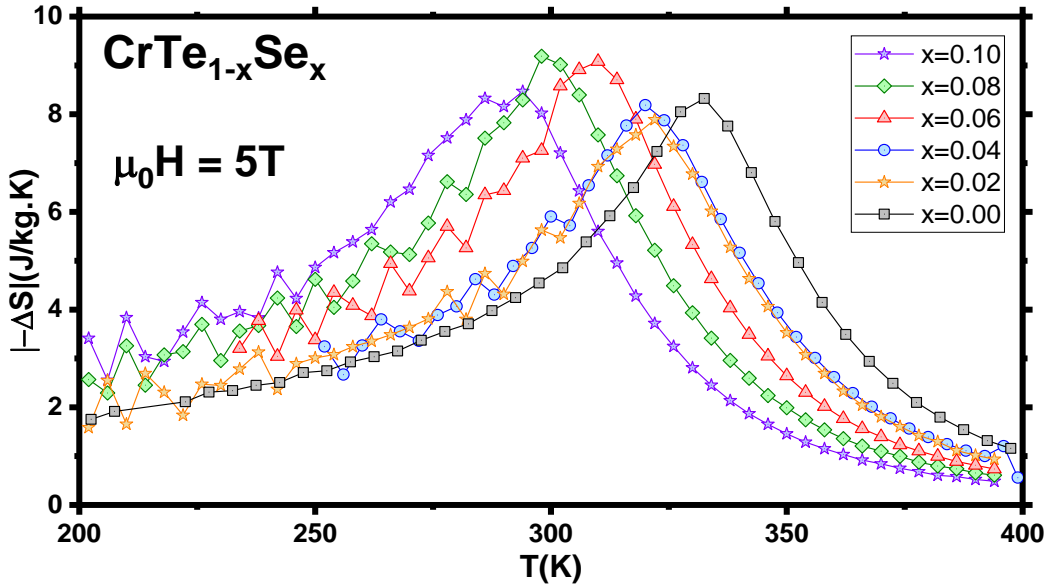

*Figure 6: Temperature dependence of $|-\Delta S(H,T)|$ of $CrTe_{1-x}Se_x$ at maximum cycling field of 5T.*

Table 4: Maximum entropy change and RCP value for $CrTe_{1-x}Se_x$ samples and different other materials for comparison at field of 5T.

| Material | Tc (K) | $|\Delta S_{max}|$ (±0.3) (J/kg.K) | $\Delta T_{ad}$ (K/T) | RCP (±0.8) (J/kg) | Ref. |
|---|---|---|---|---|---|
| Gd | 294 | 10.2 | 3 | 410 | [17, 30] |
| $Gd_5Ge_2Si_2$ | 275 | ~18.5 | | 425 | [31] |
| $Gd_5Si_{2.06}Ge_{1.94}$ | 306 | 10.81 | | 735 | [33] |
| $MnFeP_{0.45}As_{0.5}$ | 300 | 18 | | 485 | [32] |
| $La_{0.7}Ca_{0.25}Sr_{0.05}MnO_3$ | 275 | 10.5 | | 462 | [34] |
| CrTe | 332 | 8.3 | 1.5 | 555.3 | This Work |
| $CrTe_{0.98}Se_{0.02}$ | 326 | 7.9 | 1.4 | 568.0 | This Work |
| $CrTe_{0.96}Se_{0.04}$ | 318 | 8.2 | 1.5 | 546.9 | This Work |
| $CrTe_{0.94}Se_{0.06}$ | 310 | 9.1 | 1.7 | 638.1 | This Work |
| $CrTe_{0.92}Se_{0.08}$ | 299 | 9.2 | 1.7 | 694.2 | This Work |
| $CrTe_{0.90}Se_{0.10}$ | 295 | 8.5 | 1.4 | 668.1 | This Work |

**Conclusion**

Polycrystalline $CrTe_{1-x}Se_x$ (x=0.00, 0.02, 0.04, 0.06, 0.08, and 0.10) samples have been characterized by x-ray diffraction and magnetization isotherms for possible use in magnetocaloric cooling. Rietveld refinement revealed two phases in all prepared samples; a minor monoclinic phase that decreases with Se concentration until it vanishes near 10% of Se, and a major hexagonal phase. All investigated alloys $CrTe_{1-x}Se_x$ showed ferromagnetic phase transition with $T_c$ that is systematically decreased with increasing x values. The sample with x = 0.08 has $T_c$ at room temperature (~299 K), as well as the highest value of maximum magnetic entropy change of 9.2 J/kg.K at $\mu_0 H$=5T. Also, the relative cooling power was highest for x = 0.08 (694 J/kg) around Tc. The adiabatic temperature change $\Delta T_{ad}$ calculated at $\mu_0 H$= 1 Tesla is approximately 1.7 K/T for x=0.08 and 0.06 samples. This is comparing to 3K/T for the pure Gd. However, the operation range cooling (for x=0.08 and 0.06) determined by the width of $|\Delta S_M^{max}|$ are considerably larger than that in Gd. Table 4 shows the $\Delta T_{ad}$ for all studied samples.

These findings suggest that CrTe$_{1-x}$Se$_x$ (0.00≤x≤0.10) alloys are promising candidate to be used for near room temperature magnetic refrigeration applications.

**Data availability**

The datasets generated during and/or analyzed during the current study are available from the corresponding author on reasonable request.

**Conflict of interest**

The authors declare that they have no known competing financial interests or personal relationships that could have appeared to influence the work reported in this paper.

**Acknowledgement**

The authors acknowledge the support provided by the Deanship of Scientific Research at King Fahd University of Petroleum & Minerals (KFUPM), Saudi Arabia, for funding this work under project No. SB201024.

This work was, in part, performed at Ames Laboratory (I. C. Nlebedim), operated for the U.S. Department of Energy by Iowa State University of Science and Technology under Contract No. DE-AC02-07CH11358.